\documentstyle[12pt,aaspp]{article}
\def\half{{\textstyle {1\over2}}}
\def\mathrm#1{{\rm #1}}
\begin{document}
\title{A Self-Consistent Dynamical Model for the {\sl COBE}
Detected Galactic Bar}
\author{HongSheng Zhao}
\affil{Max-Planck-Institute f{\"u}r Astrophysik, 85740 Garching,
Germany\altaffilmark{1}\\
	Columbia University, Department of Astronomy, NY10027, USA\\
Email: hsz@MPA-Garching.MPG.DE}
\altaffiltext{1}{Present address}
\begin{abstract}
A 3D steady state stellar dynamical model for the Galactic bar is
constructed with 485 orbit building blocks using an extension of
Schwarzschild technique.  The weights of the orbits are assigned using
the non-negative least square method.  The model fits the density
profile of the {\sl COBE} light distribution, the observed solid body
stellar rotation curve, the fall-off of minor axis velocity dispersion
and the velocity ellipsoid at Baade's window.  We show that the model
is stable.  Maps and tables of observable velocity moments are made
for easy comparisons with observation.  The model can also be used to
set up equilibrium initial conditions for N-body simulations to study
stability.  The technique used here can be applied to interpret high
quality velocity data of external bulges/bars and galactic nuclei.

\end{abstract}

\centerline{Submitted to Monthly Notices}
\twocolumn
\tighten
\section{Introduction}

Numerous evidences have convincingly shown that our Galaxy has a
central bar with its near end on the positive Galactic longitude side
(see review by Gerhard 1995).  The interesting theoretical question at
this point is no longer proving whether such a bar exists, but rather
building a comprehensive stellar dynamical model of the bar.  The most
interesting models are those that can fit the high quality {\sl COBE}
infrared surface brightness maps of the Galaxy (Weiland et al. 1994),
which are by far the most comprehensive observational constraints on
the spatial distribution of the bar.  Such a model will also form a
basis for us to interpret the stellar velocity data, the microlensing
observations, and the gas kinematics; comparisons with these
observations require us to make inferences of the radial velocity and
the proper motions of the bar stars as well as the potential of the
bar (Zhao et al. 1994, 1995, Binney et al. 1991, Wada et al. 1994).

While there are many models for the density or the potential of the
bar both before and after the {\sl COBE} data (e.g., Blitz and Spergel
1991, Binney et al. 1991, Dwek et al. 1995), the only
stellar dynamical model of the bulge so far is an axisymmetric bulge
model (Kent 1992, Kuijken 1994).  Note that throughout the paper we
use the word self-consistent to mean that the input volume density,
which fully or partly determines the potential, is consistent with an
underlying positive distribution function.  So the potential of a
self-consistent bulge can have a bulge part and a disk part.  If there
were no disk, we say that the bulge is also self-gravitating.

Steady state 3D bar models are difficult to construct because both
integrals of motion and the distribution function have to be sought
numerically (for a general introduction, see Binney and Tremaine
1987).  The ones that also fit observations of the stellar light
and/or velocity distribution are, as far as we know are, not existent
except for Zhao (1994).  A few attempts have been made in the past
with mainly three different ideas.  One is to build a bar with a
$f(E_J)$ distribution function (e.g., the polytropic Jacobi ellipsoids
by Vandervoort 1980), where $E_J$ is the Jacobi's integral in a bar
potential.  As a result surfaces of equal density and equal effective
potential coincide in these models.  As the surfaces of equal
effective potential are too round, solutions, if exist, are only
mildly triaxial with a flat density profile, a rigorously solid-body
rotation and an isotropic velocity distribution, none of which
resemble general bars (Binney and Tremaine 1987).

The second idea is to build a system from a set of stellar orbits
without using any analytical integrals, which is known as the
Schwarzschild (1979, 1982) technique.  As box orbits are generally
more flattened and triaxial than their surfaces of equal effective
potential one can build a wide range of flattened and triaxial
systems.  The technique was successfully applied to elliptical systems
without much figure rotation by Schwarzschild and to 2D rapidly
rotating bar models by Pfenniger (1984).  Pfenniger also justified the
use of irregular orbits, and applied the non-negative least square
method (hereafter NNLS) to fit photometric and kinematic data.
However, perhaps due to many practical difficulties, the Schwarzschild
method has not been tried again in making 3D rapidly rotating bars.

Perhaps the most popular idea to build bars is the third idea, which
is to run N-body simulations of an initially unstable disk.  This idea
is straightforward and has the added advantage of giving a stable bar
as well.  However, although some models (e.g., Fux et al. 1995a,b,
Sellwood 1993) can even match observations qualitatively, the approach
do not have the build-in freedom to fit observations in detail.

This paper presents a 3D bar model that fits the {\sl COBE} maps the
Galactic bar as well as a handful of radial velocity, radial
dispersion, proper motion dispersion and metallicity measurements of K
and M giants in the bulge (see de Zeeuw 1993).  The work here is a
continuation of our efforts to link these data in a dynamical model
(Zhao et al. 1994, Zhao 1994).

The technique here is primarily Schwarzschild method combined with the
advantages of the two other methods.  Part of the bar's mass is
assigned to numerically integrated direct regular orbits, the rest of
the mass is given by a $f_c(E_J)$ distribution, which implicitly and
efficiently takes into account of the chaotic orbits without
introducing time-dependency to the model.  While the $f_c(E_J)$
component can set up a roundish system with about the right density
and velocity dispersion profile, the direct regular orbits let the
system have a boxy and barred shape and an anisotropic velocity
ellipsoid.  The stability of the model is also tested with the N-body
method.

The NNLS method is used to incorporate smoothness into the model.  As
stressed by Merritt (1993), non-uniqueness is a typical problem of
inverting distribution function based on data of its lower dimensional
projections.  We enforce our model DF to be smooth and positive so as
to lift the mathematical degeneracy of among many equivalent
(approximate) solutions and to recover a likely realistic stable bars.

The structure of the paper is as follows.  Following a brief overview
of the technical developments since Schwarzschild's pioneering work,
we lay out the modelling technique, particularly modifications to the
original Schwarzschild technique, in enough technical details so that
interested people can write their own program based on them in
\S2.  The key data used in this analysis, the {\sl COBE} Galactic
plane map and the stellar kinematic data, are given in \S3.  \S4 gives
the model results, and compares them with observations.  \S5 addresses
other important issues of the model, including the fraction of the
retrograde orbits and the chaotic orbits and their implication on bar
formation, and the uniqueness of the model.  Most importantly, the
stability of the model was tested with N-body simulations.  Finally we
summarize the technique and the results and point out a few directions
for future work in \S6.

\section{ Modelling technique }

\subsection{Overview}

Although very successful historically, the Schwarzschild technique has
not been applied to model real bars due to several practical
difficulties.  To build a stellar bar, one needs a complete
but not redundant library of time-invariant galaxy building blocks.
This already puts a lot of demand on fast computers, good algorithms
to handle force calculation and orbit integration, good understanding
of closed orbits in a 3D bar potential, good automated schemes to
launch, classify and select orbits, and large memory space to store
many orbits.  However, the main difficulties are perhaps the handling
of chaotic orbit, a plausible input model from observation, and a
robust technique to derive a smooth distribution function.

Since the pioneering works of Schwarzschild (1979, 1982) on 3D slow
bars and Pfenniger (1984) on 2D rapid bars more than a decade ago,
computers have increased their speed thousand-fold and increased swap
space and disk space greatly as well.  Simple and efficient Poisson
solver, e.g., the orthogonal basis expansion method (Zhao 1995), are
also available.  Our understanding of orbit structure in bars has also
deepened (see review by Sellwood and Wilkinson 1993 and Contopoulos
and Grosb{\o}l 1989 and Athanassoula 1992).  While isolatedly they are
not critical, together, as we will show, they make it possible to
build a large and complete orbit library.

Chaotic orbits certainly occupy most of the phase space of a 3D bar,
and make up a significant fraction of the mass of the bar as well
(Pfenniger 1984).  But the time averaged properties of an individual
chaotic orbit appear to converge only very slowly, if at all, with
increasing length of integration.  Direct computation of these orbits,
which is the approach taken by Merritt and Fridman (1995) but for
somewhat different systems, is extremely expensive.  A new efficient
method will be presented later in this section.

Observations are obviously important, because nature makes dynamically
consistent 3D bars.  For the Galactic bar, high quality {\sl
photometric data} from {\sl COBE} provides a sensible input for the
bar's density (Dwek et al. 1995) and potential, the {\sl spectroscopic
and proper motion data} constrains the velocity distribution of the
orbits (Zhao et al. 1994), and the corotation radius can be estimated
based on locating bar resonances in the {\sl gas distribution} (Binney
et al. 1991).  These informations get us immediately close to the
parameter space of a dynamically consistent bar.  How to incorporate
the variety of data will be discussed.

A {\sl robust algorithm} for the inverse problem is also crucial for
deriving a physically plausible solution among many mathematically
similar ones.  The problem can be regularized by using smooth basis
functions (Dejonghe 1987, 1889) or using the NNLS method (Merritt
1993), both of which can enforce positivity, smoothness and (to some
extent) stability of the model distribution function, and fit
observations of light and velocity.  But it needs to be worked out how
to smooth when the integrals of motions are not explicitly known.

In the following, we give the technical details and discuss some
complications of the above issues in building a 3D dynamical bar
model.

\subsection{Formulation of a stellar dynamical model with
observational constraints }

Given a set of integrals of motion, $(I_1, I_2, I_3)$, or in short,
${\bf I}$, the problem of dynamical modeling becomes finding a
realistic superposition of the orbits, or the distribution function
(DF) $f({\bf I})$, which satisfies the following four main
constraints.  The model must be self-consistent
\begin{equation}\label{selfeq}
\int d^3{\bf v} f({\bf I}) = \rho({\bf x})
\, .
\end{equation}
This equation defines what we mean by self-consistency precisely: the
input stellar mass 3D density $\rho({\bf x})$ is a projection of the
6D orbit distribution $f({\bf I})$ to the volume space.  This relation
does {\sl not directly} involve the Poisson's equation, even though
$\rho({\bf x})$ often fully or partially determines the potential
$\Phi({\bf x})$, which in turn acts on $f({\bf I})$ through the
integrals of motion ${\bf I({\bf x}, {\bf v}; \Phi({\bf x}))}$.

The model needs to fit observations, which generally can be written as
follows,
\begin{equation}\label{obseq}
\int\! dr \int\! \int\! dv_\alpha dv_\delta f({\bf I}) =
\nu(\alpha, \delta, v_r) \, ,
\end{equation}
where $\nu(\alpha, \delta, v_r)$ is an observable projected
distribution function, which equals the number of observed stars in
our Galaxy with the line-of-sight velocity $v_r$ in the direction
$(\alpha,\delta)$, or a velocity profile of unresolved stars in a sky
direction for external systems.

$f({\bf I})$ must also be positive definite,
\begin{equation}\label{poseq}
f({\bf I}) \geq 0 \, ,
\end{equation}
and be plausiblely smooth,
\begin{equation}\label{smeq}
\sqrt{\lambda} \int d{\bf I}  {\bf S} f({\bf I}) \sim 0 \, ,
\end{equation}
\noindent
where ${\bf S}$ is a linear operator, and $\lambda$ is a tunable
constant between zero and unity to steer the solution between the
wildly oscillating ones which fit the data exactly to very smooth ones
with a large residual; we expect the realistic solutions are between
the two extremes.

Deriving DF is an inverse problem which deals entirely with
deprojecting Equations~\ref{selfeq} and~\ref{obseq} with the
constraints~\ref{poseq} and~\ref{smeq}.  Except for two-integral
axisymmetric or spherical systems (Hunter and Qian 1993), inversion
has to be done numerically for a wide range of realistic systems,
including axisymmetric systems of the third integral, triaxial systems
and bars.

Also note that the observed surface brightness, the velocity profile
and its moments, as well as the volume density, are various moments of
the phase space density $f({\bf I})$.  So the Equations~\ref{selfeq}
and~\ref{obseq} can be rewritten as a more compact and general linear
equation,
\begin{equation} \label{eq:Pf}
\int d{\bf I} {\bf P} f({\bf I}) = {\bf \mu} \,
\end{equation}
where ${\bf \mu}$ denotes a certain known moment of the distribution
function, which can be observables, e.g., the surface brightness and
line profile, or known model quantities, e.g., density.  ${\bf P}$ is
a constant projection or moment operator from phase space to the
observable space.

The general form of equation~\ref{eq:Pf} is also valid to formulate
observable velocity moments as well as the line profiles.  Sometimes
the data is only good enough to give a few moments of the profile with
certainty, for example, the projected density (the zeroth moment), the
radial velocity and dispersion (the first and second velocity
moments), skewness and kurtosis (the third and the fourth moments).
In other cases, one has proper motion information, the dispersions and
the cross terms of the velocity ellipsoid as in the Galactic bulge
data analyzed in Zhao et al. (1994).  These moments can all be
programmed in a linear form similar to Equation~\ref{eq:Pf}.  For the
zeroth moment, namely, the projected density, we have
\begin{equation}\label{projm}
 \int\! dr \int\! d^3{\bf v} f({\bf I}) = \mu (\alpha, \delta)
\, .
\end{equation}
For the line-of-sight velocity moments with $n=1,2,3,4,...$, we have
\begin{equation}\label{vn}
\int\! dr \int\! d^3{\bf v} f({\bf I}) (v_r^n-\eta_n)=0\, ,
\end{equation}
where $\eta_n$, the $n$-th observed line-of-sight velocity moment, is
defined by
\begin{equation}
\eta_n \equiv \left< v_r^n \right>\, .
\end{equation}
In detail,
\begin{eqnarray}
\eta_1 &\equiv& V_r \\
\eta_2 &\equiv& \sigma_r^2+V_r^2 \\
\eta_3 &\equiv& \xi_3 \sigma_r^3 +3 \sigma_r^2 V_r +V_r^3 \\
\eta_4 &\equiv& \xi_4 \sigma_r^4 +4 \xi_3 \sigma_r^3 V_r +6
\sigma_r^2 V_r^2 +V_r^4
\end{eqnarray}
where $V_r$, $\sigma_r$, $\xi_3$ and $\xi_4$ are the observed mean
velocity, dispersion, skewness and kurtosis respectively.  One can
also program the full velocity dispersion tensor in similar way.

The moments, particularly, the skewness $\xi_3$ and the kurtosis
$\xi_4$, are often hard to obtain accurately from observations because
of the finite noise at the high velocity wings (van der Marel and
Franx 1993).  Better constrained from observation are the first few
coefficients, $h_n$ for $n=1,2,3,4,...$, of the Gaussian-Hermite
expansions of the observed velocity profile.  In this case, the
constraints become
\begin{equation}\label{hn}
\int\! dr \int\! d^3{\bf v} f({\bf I}) \alpha(w) (H_n(w)-h_n)=0
\end{equation}
for $n=0,1,2,3,4,...$, where $\alpha(w)$ and $H_n(w)$ are the same
Gaussian and Hermite functions of a renormalized velocity
$w=(v_r-V_0)/\sigma_0$.  The values of $V_0$ and $\sigma_0$ need to
specified before the NNLS process.  They could be the best-fit values
from the observed profile as in van der Marel and Franx (1993), or
some predefined value as in Gerhard (1993).

It has been argued that since the velocity dispersion is not a linear
function of DF, it can not be used as constraints in the NNLS method
(Pfenniger 1984).  But this is hardly a problem, because from the
observed dispersion $\sigma_r$ and mean velocity $V_r$ one can easily
reconstruct the second moment $\eta_2 \equiv \left< v_r^2 \right>$,
which constrains DF by a linear equation, namely Equation~\ref{vn}
with $n=2$.  Similarly for programming the skewness and kurtosis.  For
the Galactic bar, we fit the projected mass distribution according to
Equation~\ref{projm}, and the observed velocity and dispersion
according to Equation~\ref{vn} for $n=1,2$.

Also note that the various moments of DF are often coupled themselves,
e.g., the model density is related to the surface brightness by the
line-of-sight integration, and is related to the potential by the
Poisson equation.  These equations must also be satisfied for the
model to be dynamical consistent, however, they do not constrain the
distribution function directly.

\subsection{Numerical implementation}

For numerical calculations of non-integrable systems, one needs to
discretize the basic equation of the dynamical model
Equation~\ref{eq:Pf}.  The continuous variable $f({\bf I})$ is
replaced with a sum of many $\delta$-functions, each of which
represents an orbit.
\begin{equation}
f({\bf I}) =\sum_{k=1,N} w_k \delta({\bf I}-{\bf I}_k) \, ,
\end{equation}
where $w_k$ is the amount of mass, or weight, on orbit $k$, and $N$ is
the number of orbits.

The continuous observables ${\bf \mu}$ in equation~\ref{eq:Pf} is
also replaced by an array $\mu_j$ with the index $j$ from 1 to $n_c$,
the number of cells in the observable space.  The array $\mu_j$ can be
the mass in a spatial cell, the observed intensity in a sky pixel and
a radial velocity channel etc..

With these, the model reduces to finding a orbit
distribution array $w_k$ with
\begin{equation}\label{eq:wpos}
w_k \geq 0 \;\;\; \mbox{\rm for $k=1,N$,}
\end{equation}
and
\begin{equation}\label{eq:mu}
\sum_{k=1,N} B_{j,k}w_k=\mu_j  \;\;\; \mbox{\rm for $j=1, n_c$.}
\end{equation}
The equation has the same meaning as Equation 1 and 2 in Schwarzschild
(1979).  Here $B_{j,k}$ form a matrix of the contribution of the
$k$-th orbit to the $j$-th observable, which enters in the $j$-th
constraint equation.  The matrix B is effectively the discrete form of
the projection operator P in equation~\ref{eq:Pf}.  It is often a
sparse matrix as an orbit often makes a thin tube in the volume space.
The computation of $B_{j,k}$ involves integrating the k-th orbit, and
finding its projection to the $j$-th observable.  If the $j$-th
observable is the volume density or the mass in certain cell j, then
the contribution will be the fraction of time that the orbit $k$
spends in cell $j$ during the integration.  If the $j$-th observable
is a velocity moment in cell $j$, then the contribution should be
multiplied by the velocity moment when the orbit passes the cell.

\subsection{Smoothing in the effective integral space}

Often there are a multiple of exact solutions or approximate solutions
to Equation~\ref{eq:mu} and the positivity equation~\ref{eq:wpos},
which are equally good or bad in satisfying these constraints.  This
is partly because the absence of observational constraints makes
deriving DF ill-conditioned.  For the Galactic bulge, this could be
due to the missing data where the extinction is heavy, and the general
paucity of proper motions and distances of stars.  A typical effect is
that one cannot constrain the 3D volume density models uniquely based
on fitting the 2D surface brightness distribution a triaxial system
(e.g., Stark 1977, Dwek et al. 1995).  More interestingly, the phase
space density cannot be uniquely constrained from the volume density
alone due to, among other things, the trade off among loop orbits of
two senses of rotations and different thickness in the same spherical
system (Lynden-Bell 1960, Dejonghe 1987), or the same
St\"ackel model (de Zeeuw, Hunter and Schwarzschild 1987, Statler
1991, Arnold et al. 1994), or the same 2D bar (Pfenniger 1984).
Fortunately velocity measurements often can distinguish a model from
another.

Even when physical constraints are complete, non-uniqueness may arise
because of noise, finite grid size, and a mathematically unstable
algorithm.  This has been addressed by Schwarzschild (1979) and
well-discussed in Numerical Recipes (Press et al. 1992) and recently
clearly demonstrated by Merritt (1993).  In principle, one can use all
kinds of indicators to distinguish among solutions, e.g., maximum
streaming or cylindrical rotation or isotropic velocity dispersion.
But Merritt proposed smoothness as the most plausible constraint to
regularize the solution.

We also believe that smoothness has some general physical arguments
behind it.  The initial sharp features in the DF might have been
smoothed out during the violent relaxation phase of galaxy formation
as stars are scattered off to all directions by potential temporal
fluctuations or by giant molecular clouds (Lynden-Bell 1967; Spergel
and Hernquist 1992).  Also very cold systems with gaps and sharp
features in the phase space can lead to secular evolution of the
potential, e.g., the growing of a nucleus or the drifting of the bar
pattern speed.  During these processes stars can be converted from one
orbit to another, which might fill the gaps and smooth the DF and keep
the system long-lived.  But on the other hand the dynamical processes
that make the system smooth may not operate to completion.  Gaps are
ubiquitous in surface-of-sections of triaxial and bar potentials.
Systems with sharp features in the observable space, e.g., ellipticals
with kinematic detached cores or boxy bulges, can not be the
projections of a very smooth phase space distribution.

In summary smoothness constraints are generally helpful in eliminating
a vast range of unphysical solutions, but should be given only a small
weighting (i.e. the parameter $\lambda$) in the modelling to avoid
assuming too much of the phase space {\sl a priori}, particularly for
triaxial or bar systems.  $\lambda$ should be small so that the
solution is never too far away from the best fit to observations, but
non-zero so that numerical noise and unphysical sharp variations in
the DF are reduced wherever possible.

We still need to decide on an explicit functional form for the
smoothness operator ${\bf S}$.  Merritt (1993) adopted ${\bf S}$ as
some kind of second order derivative of $f({\bf I})$ in the known
integral space $(I_1, I_2, I_3)$.  Although this works very well for
systems with analytical integrals of motion, it is not very meaningful
for bars.  For our system, we find that a more applicable and
straightforward approach involves minizing the difference between the
weight on an orbit and the average weight of nearby orbits.
Mathematically, this means minimizing
\begin{equation}\label{smoothdef}
\lambda (S w)^2 = \lambda
\sum_{k=1,N} (w_k- \sum_{k'=1,n} s_{k,k'} w_{k'}/n)^2\, ,
\end{equation}
where $n$ is the number of the nearest neighbours, and $s_{k,k'}$ is a
smoothing kernel, which a function of distance between $k$ and $k'$
orbits.  To define nearby orbits, one can use the effective integrals,
which we will come to define, as indicators of proximity so that
orbits with similar energy and angular momentum are given similar
weights.  If ${\bf I}=(I_1, I_2, I_3)$ are three dimensionless
effective integrals of order unity, then $({\bf I}-{\bf I}')^2$
defines a distance between two orbits.  The kernel $s$ in
Equation~\ref{smoothdef} is chosen to be a Gaussian function of the
distance which peaks at zero.  Typically the number of nearest
neighbours $n$ is about $3^3-1=26$.  We fix the smoothness measure at
$\lambda=N^{-2}$, where $N$ is the number of orbits.  The formulation
here is certainly not the only one.  But for our model, we find
various other choices of smoothing and weighting give largely
equivalent mathematically stable results.

The lack of analytical integrals of motion other than $E_J$ for a
general bar potential opens up a variety of effective integrals to
serve as descriptions of an orbit.  If $E$, $J_z$ and $J_x$ are an
orbit's instantaneous energy and angular momentum components in the
short $z$ axis and in the long $x$ axis, then the time averaged
quantities $\langle E\rangle$, $\langle J_z\rangle$ and $\langle
J_x^2\rangle$ can be used to describe the most important properties of
the orbit, namely, its radial extent, its sense of rotation along the
minor axis and its vertical extent.  Since ($\langle E\rangle$,
$\langle J_z\rangle$, $\langle J_x^2\rangle$) reduce to exact
integrals in oblate or prolate potentials, which are two extremes of a
bar potential, they seem to be better choice of the effective
integrals.  Other quantities, such as the axis ratio ($y/x$, $z/R$)
defined by square root of ratios of the time averaged principal axes
of moment of inertia tensor for an orbit, are also useful effective
integrals.  We rescale all these effective integrals to the range of 0
to 1, and define the nearness between orbits used in the smoothing
equation~\ref{smoothdef}.

\subsection{NNLS}

Equation~\ref{eq:mu}, ~\ref{smoothdef} and~\ref{eq:wpos} together
define a NNLS problem.  The dynamical model now becomes a
solution that minimizes the following $\chi^2$.
\begin{equation}\label{eq:chi2}
\chi^2 =  \lambda (Sw)^2 + \sum_{j=1,n_c} (\sum_{k=1,N}
B_{j,k}w_k -\mu_j)^2/\sigma^2_j \, .
\end{equation}
The first term on the right hand side is the smoothness.  The second
term is the sum of residuals in fitting the observables ${\bf \mu}$.
Each observational constraint is made dimensionless by a scaling
quantity $\sigma_j$, which we set to be the rms value of the
observable $\mu_j$.  The goodness of a fit will be measured by a
dimensionless residual in the fit to the density, the projected
density and the velocity.  Minimizing the above $\chi^2$ yields a
smooth solution among those that fit observations.

The standard algorithms to solve a NNLS problem can be found in QPROG
of IMSL and E04NAF of NAG.  Numerical Recipes (Press et al. 1992) also
gives an excellent discussions of the inverse problem and NNLS method.
For the computation here, we use a software package in the public
domain of AT\&T, which is available through anonymous ftp.  The main
subroutine is named ``dwnnls.f''.  A similar source code is also
available in Hanson and Lawson (1974).

\subsection{Time-dependency, completeness and collective-orbits}

The use of numerical orbits in making steady state models introduces
some worrysome problems which concern less models with analytical
integrals.  An orbit model with irregular orbits could be
time-dependent and very sensitive to local perturbations.  Only the
regular orbits with, by definition, three integrals of motion are
legitimate building blocks for a steady state model.  Regular orbits
with nearly the same integrals librate close to each other and around
a stable periodical orbit of the bar.  Their properties as described
by the effective integrals, e.g., $(\langle E\rangle, \langle
J_z\rangle, \langle J_x^2\rangle)$, depend little on the length of
integration after some ten rotations and are insensitive to small
changes in initial conditions.

On the other hand, irregular orbits may be important in real bars.
For example, an orbit model with regular orbits only has a peanut
shape in 3D, because the regular orbits in a rapidly rotating bar
potential typically have a fixed sense of rotation, and the
centrifugal force prevents them from reaching the minor axis (Zhao et
al. 1994).  But the Galactic bar model is probably not peanut-shaped,
as in the {\sl COBE} map and the volume density model by Dwek et
al. (1995).

The irregular orbits have finite time-dependency even after
integrating much longer than their orbital time scales.  Pfenniger
(1984) finds that the spatial density of an irregular orbit averaged
over multiples of Hubble time fluctuates on the level of a few
percent.  We also find that irregular orbits seem to come in two
kinds, semi-regular and chaotic.  While a chaotic orbit occupies a
featureless volume bound by its Jacobi's integral in the outside and
some islands of regular orbits in the inside, a semi-regular orbit
keeps some structures for at least 100 rounds of the Galaxy, e.g., it
can keep its sense of rotation around the Galaxy and a hole in the
center.  Semi-regular orbits and chaotic orbits can exist at the same
energy with regular orbits (see the surface of section plots of Figure
13 in Sellwood and Wilkinson 1993).  It is possible that a
semi-regular orbit and a chaotic orbit with the same Jacobi integral
are two long phases (much longer than a Hubble time) of one orbit,
which happen to be very close or very far from the regular islands.
There may be on the order of $10^6-10^{10}$ stars on the same
irregular orbit, spread out in the orbital phase.  Treating the
semi-regular phase and the chaotic phase as independent orbits also
causes the model to evolve with time, but often on time scales too
long to have any effect on steady state models.

The three types of orbits also have different levels of
time-dependency.  We find that while a chaotic orbit fluctuates on
typically 10\% level as the integration is not long enough to fill its
5D space evenly, a semi-regular orbit and a regular fluctuate less
than 5\% and 1\% respectively as the phase space is roughly 3D.  Only
regular orbits are shown in the lower panel of
Figure~\ref{orbit-point.ps}.

As in a real galaxy, small level of potential fluctuation and secular
evolution are common, some small amount of time-dependency should also
be allowed for the models.  Bars are typically young, less than 100
rotation periods, so even a regular orbit still has some memory of its
initial phase, and the system may only be marginally in steady state.
The secular effects of irregular orbits on a model may well be
negligible as far as fitting observation is concerned.  One also
expects that some irregular orbits have been populated by relaxation
processes during its formation.  As they also contribute to the
observed light and velocity, it is natural and necessary to include
them in our orbit model so as to be in equilibrium and to be
consistent with observation.

To achieve the completeness without introducing the time-dependency
problem of chaotic and semi-regular orbits, one needs an alternative
type of building blocks other than numerical orbits.  Models built
with analytical integrals offer some hints.  In isotropic spherical
systems or axisymmetric systems, one can build models analytically
with distribution functions $f(E)$ or $f(E,L_z)$.  Likewise, one can
build isotropic bar models, namely Jacobi ellipsoids, with a
distribution function $f(E_J)$ based on the only known integral $E_J$.
A distribution function $\delta(E_J-E_{J0})$ prescribes a group orbits
with the same $E_J=E_{J0}$ populated with equal weight.  We call such
a constituent of the bar a ``collective-orbit''.  A model with a DF
$f(E_J)$, namely, a superposition of collective-orbits, is known to be
either non-self-gravitating or far from real stellar systems (see
Introduction).  But a distribution function $f_c(E_J)$ can prescribe
``part'' of the mass in a bar; the rest, $f_n$, can still be from the
numerically computed orbits, so that the system's distribution
function $f(\bf{I})=f_n(\bf{I})+f_c(E_J)$.

Collective-orbits are legitimate constituents of any equilibrium
models, just the same as numerically computed regular orbits, because
both are solutions of the Vlasov equation $\frac{d}{dt}f=0$.  A
collective-orbit occupies a 5D volume, which may include some regular
islands but mostly chaotic regions.

Although they occupy a similar 5D region in phase space as chaotic
orbits, collective-orbits are free from the large fluctuations,
which may or may not be beaten down completely by carrying out a high
precision integration to a formidably long time scale.  The phase
space distribution of a collective-orbit is analytical.  To know their
property in the observable space involves no orbit integration, only a
few trivial projections.  These nice properties of collective-orbits
and the time-dependency problem of chaotic orbits are also
well-demonstrated in a recent detailed study of orbits in triaxial
cusped nuclei by Merritt and Fridman (1995).

We explicitly use collective-orbits as galaxy building blocks in our
model.  We use them to replace the hard-to-handle semi-regular orbits
and chaotic orbits.  Using collective-orbits also makes our model a
hybrid model of he ones with analytical integrals and the ones with
only numerical orbits.  Both are critical in matching observations.
In actual implementation, the functions $f_n(\bf{I})$ and $f_c(E_J)$
are implied functions.  Only the weights to the orbits $w_k$ enter the
calculation, which are determined by fitting observations.

\subsection{Constructing the orbit library}\label{lib}

A complete orbit library is essential in building a steady state
model.  The orbit structure of a bar potential is complex with large
numbers of resonances and bifurcations.  The regular and the
semi-regular $x_1$ and 2:2:1 orbits, which are the backbone of the
bar, must be well represented in the orbit library.  On the other
hand, the retrograde orbits and the chaotic orbits occupy most of the
phase space, and they can be important in short fat bulge-like bars
without peanut shape and strong direct streaming motion, which may be
the case for the Galactic bar (Zhao et al. 1994).  Besides these major
orbit families, the numerous minor orbit families may well play some
role in filling the gaps between major families and contribute a
smooth density and velocity distribution.

Previous workers, who were limited by the speed of their computers,
could only afford to populate the main orbit families, which can be
reached by launching orbits perpendicularly from the intermediate axis
of a bar (Schwarzschild 1979, 1982; Pfenniger 1984).  This approach
is effective in finding regular orbits, but it could miss some orbits
that also occur in the steady state bar.  Alternatively one can
populate the orbits in a Monte-Carlo fashion, and launch orbits with
random initial conditions to cover all possible initial conditions
(Zhao et al. 1994, Zhao 1994).  As there are large stochastic regions
in a rapidly rotating bar plus a nucleus, the uniformly launched
orbits would bias strongly towards the chaotic orbits, leaving very
few regular orbits in a finite orbit library.  The lack of boxyness in
Zhao's (1994) model is mostly due to this effect.

We control the initial orbit distribution in the following way.  Each
orbit is launched tangentially with a speed less than the circular
velocity from a local apgalacticon.  The unspecified coordinates are
chosen randomly but with minimal clustering to reduce redundancy.  The
initial radius is sampled uniformly between zero to $4$ kpc.  Most
orbits are launched in close pairs perpendicularly from the $xz$, $yz$
or $xy$ symmetry plane or the $x$ or $y$ symmetry axis, and a small
fraction of orbits are either from the minor axis or from no
particular positions.  Collective-orbits are included in our
library; for them the computation involves only projections, no orbit
integration.  For the rest, we stop the integration of an orbit when
it has made more than 100 radial oscillations, which is roughly 50-100
rotation periods; every two successive radial turn-backs are counted
as one oscillation, or one epicycle.  During the integration of the
orbits, we keep track of the energy, angular momentum and the axis
ratio of the orbit so that its sense of rotation and shape are
determined.  Some orbits have the tendency to escape from the bulge.
They either have a positive instantaneous energy or reach beyond 7 kpc
at one time.  We also trace the deviation between the orbit pairs to
classify an orbit as regular, semi-regular or chaotic.  At the end of
an orbit integration, a regular orbit would typically has filled its
3D torus in phase space.  In fact, we call an orbit regular only if
its pair orbits never diverge more than linearly and the fluctuation
of the effective integrals is less than 1\% (see
Figure~\ref{orbit-point.ps}).  The chaotic orbits, on the other hand,
fluctuate at typically 10\% level at the time we stop the integration.
Collective-orbits have intrinsicly zero fluctuations.  The
fluctuation here is the difference in the time averaged moments of an
orbit for integration length $t/2$ and $t$, where $t$ is about 100
orbital rotations.

Not all of these orbits are kept for the final model.  From many
preliminary runs, we come to realize that the most important orbits in
making the bar are the collective-orbits and the direct boxy regular
orbits.  Since direct regular orbits are intrinsically rare in the
phase space, most of the phase space is taken by retrograde orbits,
chaotic orbits and escaping orbits.  The orbits from the minor axis or
some random positions often pass the central nucleus in a few
dynamical times, and inevitably end up as chaotic orbits.  To best
make use of our computer resources, we discard ecaping orbits, chaotic
orbits, semi-regular orbits and retrograde orbits.  These orbits are
only implicitly contained in collective-orbits.

{\sl Only} the two critical families, namely, the collective-orbits
and the direct regular orbits, are carried into the final modelling
process.  Both have the nice property of being time-independent.
While the former makes up a roundish bulge with certain density
profile along the minor axis, the later adds to it barred and boxy
features and velocity anisotropy.  Only together they form a system
close to the Galactic bar.

\subsection {Orbit integration details}\label{orbit_detail}

The orbits are integrated using a Bulirsch-Stoer integrator (Press et
al. 1992) with large uneven time steps, and the intermediate position,
velocity and acceleration are stored as unformatted data file on disk.
The task can be spread out among several processors computing
different orbits at the same time.  One processor collects and
processes these data files off-line.  The data are interpolated with
high order polynomials and then binned according to the position,
velocity, sky direction, or line-of-sight velocity to obtain the final
information of the projected spatial and velocity distribution of the
orbits.  The interpolating polynomial is fifth order for the position
coordinates, four order for the velocity and third order for the
acceleration.

Comparing with other ways, e.g., a direct integration of the moments
along with orbits in fine steps, this procedure has several
advantages.  Our integrator is much more efficient than the Runga-Kuta
integrator for smooth potential and is able to make large time steps
without reducing accuracy.  Since the time steps are large, one can
afford to store all the orbits.  The piecewise high order polynomials
interpolate within a big step accurately without propagating errors
from steps to steps.  Paralell processing through disk also makes best
use of resources.

Storage space can be a limiting factor of the size of the simulation.
To avoid recomputing the projections of the orbits in the NNLS part,
we need to store at least the spatial density in a 3D grid and the
projected density and radial velocity first two moments in a 2D grid
for each orbit.  Since the orbital equations in a bar potential
preserve the reflection symmetries $(z \rightarrow -z$, $V_z
\rightarrow -V_z)$, $(y \rightarrow -y, V_x \rightarrow -V_x)$, and
$(x \rightarrow -x, V_y \rightarrow -V_y)$, one can construct 16
mirror images for each orbit of the bar; the 16 orbits are sometimes
degenerate to each other.  We always equally populate the 16 orbits in
the model with the consideration that a steady state model is
unlikely to crucially depend on any minor axis rotation or any $m=1$
or $m=2$ spiral arm modes.  We also make use of the reflection
symmetries in storing the intrinsic and projected mass distribution of
the orbits.  A $10 \times 10 \times 10$ rectangular grid is set up to
model the volume density of the bar in the first octant with the cell
size in the x, y and z directions being $200$, $150$ and $100$ pc.
Projected maps of density, flux and pressure moments are made with one
square degree resolution in longitude $l$ and latitude $b$ within $0
\leq b \leq 10$ and $-16 \leq l \leq 16$ degrees.

It takes typically 1 minute of CPU time on an IBM RS/6000 workstation
(with computing power roughly equivalent to a Sparc 10) to select and
integrate one orbit.  One needs about 10 hours of CPU and 0.1 GB disk
space to compute and store every 1000 orbits.  The NNLS calculation
takes a modest amount of CPU time (typically one or two hours), but
needs a very large work space.  To program the $n_c$ constraints and
$n_p$ smoothness equations in double precision, one needs roughly $10
{N+n_c \over 1000} {N \over 1000}$ MB work space, where $N$ is the
number of orbits.  Typically, one can run the code with $N \sim
1000-3000$ on Sparc stations with medium swap space.

As the algorithm and computation in constructing equilibrium
models is complicated, one needs to test the procedure extensively.
We have made two ``full system'' tests with the Hernquist (1990)
spherical bulge model and the axisymmetric isotropic model of Kent
(1992).  The former model is known analytically, and the latter model
can be constructed by solving the Jeans equation.  In the tests of both
models, we do not make explicit use of the spherical or axial
symmetry.  The details of these test runs can be found in Zhao (1994).
We find that the models recover the known solutions to good accuracy,
and the basic technique is suited to constructing numerical
equilibrium models.  We now proceed to the bar model.

\section{Inputs of the model}

\subsection{ Density model of the bar}\label{dweksec}

In additional to its better known cosmological achievements, the {\sl
COBE} satellite also provided an invaluable photometric dataset for
the study of galactic structure as a result of the Diffuse Infrared
Background Experiment (DIRBE) on board (Boggess et al.  1992).  {\sl
DIRBE} has mapped the Galactic bulge within $|l|<30^o$ and $|b|<15^o$
and the Galactic plane within $|b|<10^o$ with $0^o.7\times 0^o.7$
resolution in 10 infrared bands, where the extinction is a less
serious problem than in optical.  Arendt et al.  (1994) extracted a
reddening spectrum and a extinction map by assuming the pixel to pixel
color variations in the four infrared maps are entirely due to
reddening by dust.  After correcting for the dust Weiland et al.  (1994)
presented maps of the high latitude ($|b|>3^o$) bulge region at four
infrared wavelengths 1.25, 2.2, 3.5, 4.9 $\mu m$, which clearly shows
a flattened peanut shape bulge with axis ratio $\sim 0.6$ and the
asymmetry in light distribution that is qualitatively consistent with
a Galactic bar with its near end in the first Galactic quadrant.

Dwek et al. (1995) fit a set of photometric bar models to the {\sl
COBE} map within $3^o<|b|<10^o$ and $|l|<20^o$.  This region
effectively excludes the disk and the low latitude region where
extinction is high.  One of their best fitting models is their G2
model, which is a boxy Gaussian model.  Assuming a constant
mass-to-light ratio ($\Upsilon={M \over L}$), the density of the bar
is fit by the following mathematical form,
\begin{equation}
	\rho_{G2}(x,y,z) = \rho_0 \exp\left(-{s_b^2 \over 2}\right)
\end{equation}
where
\[
s_b^4 =\left[\left({x \over a}\right)^2 + \left({y \over b}\right)^2
\right]^2 +  \left({z \over c}\right)^4
\]
The least square fit yields the scale lengths $a = 1.49 \pm 0.05$, $b
= 0.58 \pm 0.01$ and $0.40 \pm 0.01$ kpc for galactocentric distance
$R_0=8$kpc.  The long axis of the bar, the $x$ axis, points parallelly
to the direction of $l = -13.4^o$ and $b=0^o$.

This density model is only a parametrized fit to the light of the
bulge outside of a few degrees, and has not been fit to observations
within $r \sim 3^o\sim 400$pc.  The G2 model has a finite core, which
is not consistent with observations that show that the galactic bulge
has a nucleus with a steep power law $\rho(r) \sim r^{-1.85}$ (Becklin
and Neugebauer 1968) that can be extrapolated to the inner bulge
region (Matsumoto et al. 1982, Sellwood and Sanders 1988).  Kent
(1992) also finds that the light of the nucleus can be smoothly joined
with the bulge seen in the 2 micron map from IRT.  Such a nuclear
component has observable effect on the kinematics of the bulge: the
turn-over of radial velocity dispersion at a few degrees in the Kent
(1992) bulge model is an example.  Theoretically, including such a
nucleus makes it more difficult to find orbits supporting the bar.  It
can make steady state bars impossible, or at least, greatly limit
the solution space.

We include an axisymmetric nucleus in our dynamical model of the bar,
which is both required by observation and is a strong test case of the
model technique.  We use an axisymmetric nucleus for the lack of
strong evidence for a corotating triaxial component in the very center
and because the boxlet orbits may not be sufficient to support strong
triaxiality.  The model density is continuous at the transition region
and is given by the following form,
\begin{equation}
\rho(x,y,z) = \rho_0 \left[ \exp\left(-{s_b^2 \over 2}\right)
+ s_a^{-1.85} \ \exp(-s_a) \right]
\end{equation}
where
\[
s_a^2 = { q_a^2(x^2+y^2) + z^2 \over c^2}
\]
and $q_a=0.6$.  The density is also truncated beyond 3 kpc.  The
reason for the truncation is both because the Dwek et al. model is
unconstrained beyond $10^o \sim 1.4$ kpc, and that an elongated bar
ends before corotation, which is at 3.3 kpc in our model.  The
included nucleus is similar to Kent (1992).  The constant $\rho_0$
will be determined by normalizing the total mass of the bar to fit the
velocity dispersion at Baade's window.

Fig.~\ref{dered.ps} compares the density model with the {\sl COBE}
map.  It shows the projected density of the model and a dereddened
{\sl COBE} map.  A Miyamoto-Nagai disk as described below is also
included in the model.  The smoothness of the dereddened contours
shows that dust subtraction is sufficient for the bulge region.  One
can see that the modified model matches the {\sl COBE} K band map
within 10 degrees of the Galactic Center to a similar accuracy as the
Dwek et al. model.  Particularly the model matches the boxyness
and the longitude asymmetry in the dereddened {\sl COBE} map
reasonably well.  We shall later on call our density model of the bar
the modified Dwek model.  Although it is still boxy, the modified
model is significantly rounder in the $xy$ plane than the G2 model as
a result of the added axisymmetric component, and it fits the {\sl
COBE} map roughly as good.  The effective axis ratio is about
$1:0.6:0.4$.

\subsection{Potential models of the bar and the disk}

The model potential is a sum of two components, a bar and a disk.
The isothermal halo is neglected because we are interested in dynamics
in the inner 2 kpc, where a halo with reasonably large core radius
contributes little.  The bar density is based on the Dwek et
al. G2-model, but is modified to account for the inner nucleus.

The disk potential is modelled as a Miyamoto-Nagai (MN) disk with an
analytical potential of the following form,
\begin{equation} \label{eq:mn}
\Phi(x, y, z)= -\frac{GM_d}{D},
\end{equation}
where
\begin{equation} \label{eq:mn2}
D^2=x^2+y^2+(a_{MN}+(z^2+b_{MN}^2)^{\half})^2,
\end{equation}
$a_{MN}=6.5$ kpc, $b_{MN}=0.26$ kpc and total disk mass $M_d=8
M_{bar}$.  The disk parameters are chosen to have a vertical height of
0.2 kpc and together with the bar produces a flat rotation curve up to
3 kpc.  Such a disk is only a very rough approximation to the
conventional double exponential disk. It does not fit the {\sl COBE}
map in region outside 10 degrees of the center in any detail (see
Figure~\ref{dered.ps}).  However, since our primary interest is the
bulge, which is roughly self-gravitating anyway, a simple
parametrization of the disk potential is acceptable and useful.  The
analytical MN disk potential helps to increase the speed of the orbit
integration.  Scaling the disk mass with the bar mass and ignoring the
halo potential simplifies our fitting procedure, as the bar mass is
scaled out of most of the calculations, and can be obtained at the end
by renormalizing with the velocity dispersion at Baade's window.

Even with a fast computer, a good algorithm to compute the
gravitational acceleration is needed to integrate the orbits
efficiently.  Recently Hernquist and Ostriker (1992) (HO) and Zhao
(1995) showed that for force calculation in steady state galaxy models
can be very efficient if both the potential and the density are
expanded on a set of simple orthogonal basis of potential-density
pairs with the lowest order term corresponding to some simple
spherical models.

To compute the bulge potential, we have used the HO expansion
technique.  We choose the HO expansions with the scale length $a=1$
kpc.  We compute the expansion coefficients for the potential by a
Monte-Carlo integration using about one million random particles
spread over the bar.

Each expansion term is denoted with three quantum numbers $(n, l, m)$.
For a triaxial model, only the even quantum number terms are non-zero.
The expansion terms can be ordered according to the value of
$N_{nlm}$,
\begin{equation}
 N_{nlm} = 1+{m \over 2} +{l \over 4} ({l \over 2}+1) +{n_l (n_l+1)
(n_l+2) \over 6} \, ,
\end{equation}
where $n_l=n+{l \over 2}.$ For increased efficiency of orbit
integration, only the leading ten terms of the expansion coefficients
are used to construct the bar potential.

We check the effect on accuracy due to the truncation in the HO
expansion by comparing the model's circular rotation velocity curve
with the truncation set after the first 10 and 84 terms respectively.
Figure ~\ref{rotcurve.ps} plots the circular velocity along the
intermediate axis of the bar.  A bar mass of $2\times 10^{10}M_\odot$
is used.  The difference due to truncations is less than 20 km/s for
the region between 0.3-3 kpc.  Also note that both rotation curve is
in agreement with the observed flat rotation curve within 3 kpc
(Clemens 1985), although a more rigorous comparison involves computing
the velocity of the closed orbits of the bar rather than the circular
velocity (Binney et al. 1991).  Overall, the model potential appears
to be reasonable for the inner bulge.

\subsection{Kinematic data}\label{kin}

The low extinction fields of the galactic bulge, particularly Baade's
window (BW) $(l,b)=(1^o,-4^o)$, have been the target of many
spectroscopic studies.  Among the largest kinematic and abundance
samples at Baade's window are the radial velocities of 300 M giants by
Sharples et al.  (1991), proper motions of 400 K and M giants by
Spaenhauer et al.  (1992), and the published kinematics and
metallicities of 88 K giants by Rich (1988, 1990).  Of these, there
are 62 K giants for which metallicity, radial velocity and proper
motions are all measured (Zhao et al. 1994).  A larger overlap sample
has also been obtained recently by Terndrup et al. (1995a).

In addition to Baade's window, kinematics of various stellar
populations are obtained at several fields on the minor axis (see data
complied in Kent 1992).  Radial velocity distributions are also
obtained at off-axis fields about $2^o$ from the center by Blum et
al. (1994, 1995), $10^o-14^o$ by Minniti et al. (1992) and Morrison
and Harding (1992) and further out by Ibata and Gilmore (1995).

The second type of kinematic data is the radial velocity survey of the
whole bulge or regions of it.  Bulge tracers like Miras, SiO maser
stars, OH/IR stars, planetary nebulae have a roughly solid body
rotation curve with a slope about 80 km/s/kpc (see de Zeeuw 1993).
The OH/IR stars also has a nuclear component, which rotates much
faster with a slope 10 times steeper than the bulge K and M giants
(Lindqvist et al. 1992a, b).  Since it is unclear whether these stars
trace the bar as the K and M giants do (Dejonghe 1993), we do not use
these stars to constrain our model except for the solid body
rotation of the bulge.

To constrain our bar models, we use the following kinematic data: the
radial velocity dispersion in Baade's window, and the proper motion
data of Spaenhauer et al., the radial velocity and dispersion at
Minniti's $(8^o, 7^o)$ field, Blum's $(-1^o, 2^o)$ field, the overall
solid body rotation curve of slope 80 km/s/kpc; the Galactocentric
distance $R_0$ is set at $8$ kpc.  An average line-of-sight dispersion
of $113 \pm 6$ km/s for all stars at Baade's window is used to
normalize the bar's mass; the number is mostly based on the 200 M
giants from Sharples et al..  The proper motion and rotation data
helps to constrain the amount of anisotropy in the model.  Although
other kinematic data are not used as model constraints, they serve as
reference values to compare with our model.  In particular, we compare
our model with the observed minor axis drop-off of velocity dispersion
for the M giants (Terndrup et al. 1995b).

\section{Results}\label{result}

We have undertaken to build dynamical bar models that are consistent
with the modified Dwek density model.  We fix the angle of the bar at
an often-quoted value $20^o$, close to the value $13.4^o$ found by
Dwek et al..  We set the pattern speed at $60\xi^{1/2}$ km/s/kpc,
where $\xi=M_{bar}/(2\times 10^{10}M_\odot)$.  This corresponds to a
corotation of 3.3 kpc.  The pattern speed here is slightly smaller
than used by Binney et al. (1991) if the bar mass is $1-2\times
10^{10}M_\odot$.

The orbit library consists of 1000 orbits, which are piped into the
NNLS routines to fit 1000 constraints from self-consistency, and some
500 constraints from the projected density and velocity.  Note that
the projected densities and the volume density are not completely
redundant due to different boundary and grid.  Although there are
somewhat more constraints than there are unknowns and the problem
appears over-determined, some level of degeneracy among nearby orbits
due to a finite grid may still exist.  To soften the problem of
non-uniqueness and to obtain mathematically stable solutions, we
require the orbital space to be relatively smooth.  As a result of
fitting these constraints, the NNLS routine assigns 325 regular direct
boxy orbits and 160 collective-orbits with non-zero unequal weights.
The rest of the orbits have zero weight.  The 485 orbits with non-zero
weight form our best fit model.

Let us first examine the extent of self-consistency of the model.
Figure~\ref{den.ps} shows the the volume density slices in the xy and
yz plane for the orbits and the Dwek et al. model.  The differences
between the two densities in quadrature sum is relatively small $\sim$
0.5\%.  The orbits fit the density profile along the minor axis, the
elongated bar shape in the xy plane as well as the boxy contours in
the yz plane from 100 pc to 1 kpc on the major axis.  Beyond 2 kpc on
the major axis, the model contours are somewhat flatter and less
barred than the Dwek et al. model.  The model has the same
shell-averaged radial profile as the Dwek et al. model plus a nucleus.
Self-consistency can also be examined in the potential.  In terms of
the expansion coefficients, the Dwek et al. model and that of the
orbits differ by $\sim$ 0.1\% in quadrature sum.  We conclude that the
model is self-consistent.

To compare with observation, we also show the observables projected
onto the sky plane.  The upper panel of Figure~\ref{surf.ps} compares
the projected density of the orbits with that of our input model.  The
agreements in the boxyness and asymmetry are good, considering
also that the Dwek et al. model is not constrained by the {\sl COBE}
map beyond $10^o$.  The residual is 0.3\% in quadrature sum.  The
model is also compared with the {\sl COBE} map in
Figure~\ref{dered.ps} after adding a disk.  The dynamical model
fits the Dwek et al. model in projected density as well as
the {\sl COBE} map in the range of $|l|<10^o$ and $|b|<6^o$.

More interesting predictions are the rotation field and dispersion
field shown in the middle two panels of Figure~\ref{surf.ps}.  Both
maps are smooth and regular.  The mass of the bar, $(2.2
\pm 0.2)\times 10^{10} M_\odot$, is normalized by Baade's Window
dispersion $113\pm 6$km/s.  The rotation field is shown to have nearly
evenly spaced contours and indicative of a solid body rotation field
with a mean slope of $(100 \pm 10)$ km/s/kpc.  This is somewhat faster
than the observed rotation rate of various bulge tracers.  The model
velocity dispersion declines away from the center.  A detailed
prediction is given in Table 1.

To inspect the model's predictions on the dispersion more closely, we
also plot the velocity dispersion along the minor axis and other
slices of the bulge for the model and observations in Figure
{}~\ref{minormajor.ps}.  The model fits the minor axis data, including
the dispersion at Baade's window and the general trend of drop-off
along the minor axis.  The data points are from M giants at Baade's
window (Sharples et al. 1991) and several other fields on the minor
axis (Terndrup et al. 1995b), and one data point representing a
typical 110 km/s dispersion at 100 pc for the central OH/IR stars
(Lindqvist 1990a, b).  Note that we did not impose the fall off along
the minor axis as a constraint to our bar model just as in the oblate
rotator model of Kent (1992).  It appears to be an inevitable
prediction of models without strong intrinsic velocity anisotropies.
The upper panel of Figure ~\ref{minormajor.ps} also shows the run of
the velocities along the $b=-4^o$ longitude slice, and the $b=-7^o$
slice.  These two slices and their positive counterparts set the
boundaries of observable low-extinction region where the bulge light
still dominates the disk light.  This intermediate range has been the
target for velocity surveys of the bulge (Izumiura et al. 1995).  In
addition to the nearly cylindrical rotation of the bulge, the model
shows almost no dependence of velocity dispersion on longitude.  For
the central field observed by Blum et al. ($-1^o$, $2^o$), the model
predicts $V_r=-40$ km/s and $\sigma_r=128$ km/s, while the observed
values are $-75\pm 24$ and $127\pm 17$ km/s.  At Minniti's field
($8^o$ ,$7^o$), the model predicts $V_r=80$ km/s and $\sigma_r=57$
km/s, while the observed values are $45\pm 10$ and $85\pm 7$ km/s.
  At Morrison and Harding's field ($-10^o$, $-10^o$), the model
predicts $V_r=-89$ km/s and $\sigma_r=91$ km/s, while the observed
values are $-82\pm 8$ and $67\pm 6$ km/s.  The model is not able to
make reliable predictions beyond $10^o$ from the center, where the
surface density of the bar is low and better disk and halo models are
necessary.  For radial velocity predictions in other fields inside
$10^o$, see Table 1.  Overall, the model fits the observed
line-of-sight velocities and dispersions reasonably well, except that
it predicts too much rotation for the bulge at large radius.  It is
likely that too many direct boxy orbits are used to fit the boxy bar
to large radii.  This may not be a big problem because we did not
explicitly use any retrograde orbits, and because the observed stellar
populations in the bulge do not exactly follow a ubiquitous solid-body
rotation law (Izumiura et al. 1995, Minniti et al. 1992, Lindqvist et
al. 1992a), and the Dwek et al. model may have oversimplified the
shape of the bar in the $xy$ plane near the corotation.

The model also shows several characteristic signatures of bars in the
proper motions.  Bars have anisotropic velocity ellipsoids.  If
$\sigma_l$ and $\sigma_b$ are the proper motion dispersions for the
Galactic bar integrated over a line of sight, one expects that
$\sigma_l>\sigma_r$ and $\sigma_l>\sigma_b$ due to both the intrinsic
anisotropy and rotation broadening in the $l$-direction.  One also
expects that the cross term of the velocity ellipsoid $\sigma_{lr}\neq
0$ due to triaxiality (Zhao et al. 1994), where $\sigma_{lr}
\equiv {\mathrm sign}(u)\sqrt{u}$, and
$u=\left<v_lv_r\right>-\left<v_l\right>\left<v_r\right>$; the sign of
the cross term tells the orientation of the velocity ellipsoid in
the $v_l$ vs $v_r$ plane.

These signatures are clearly seen in our bar model.  Table 1 and 2
give detailed predictions of the velocity ellipsoid for fields within
$10^o$ of the center.  The lower panel of Figure ~\ref{minormajor.ps}
also shows the four moments of the velocity ellipsoid along the minor
axis.  $\sigma_l$ is clearly systematically larger than $\sigma_b$ and
$\sigma_r$, and the cross term $\sigma_{lr} \neq 0$ for almost the
entire minor axis.  At Baade's window, the model predicts
$\sigma_l/\sigma_b = (1.3 \pm 0.1)$, larger than Spaenhauer et al.'s
(1992) observation of $\sigma_l/\sigma_b=(1.15 \pm 0.06)$ for all the
K and M giants, but consistent with that of their metal rich
subsample.  On the other hand, the vertex deviation shown by the cross
term is a more definitive means of showing the triaxiality of the
bulge.  At Baade's window, we predict
$\sigma_{rl}/(\sigma_r\sigma_l)^{1/2}=-0.4$.  The result confirms the
vertex deviation seen in a small overlap sample with complete velocity
information and in the previous semi-consistent model (Zhao et
al. 1994), and strengthens the argument that the metal rich bulge is
triaxial.

In terms of planning future observations, we find the combined proper
motion and radial velocity data is more sensitive to the triaxiality of
the bar.  The observable velocity moments do not obey reflection
symmetry with respect to the $l=0$ axis due to perspective effects of
the bar (see Table 1 and 2).  But the typical difference of only 20
km/s in velocity $V_r$ and dispersions $\sigma_r$, $\sigma_l$ and
$\sigma_b$ can be difficult to observe, given typical sample size of
$100$ stars per field.  On the other hand, our model predicts that
$\sigma_{lr} < 0 $ on the minor axis and it will change sign (become
positive) for fields at negative longitude and high latitude.  As it
is easier to distinguish between two perpendicular velocity
ellipsoids, it may be worth the efforts to measure radial velocities
of a proper motion sample with about 100 stars on the minor axis or
two proper motion samples at opposite longitude fields of the bulge.

In summary our bar model fits the observations of the Galactic bulge
in the light distribution and in the kinematics and is
self-consistent.  The basic technique works very well.

As the modelling process is relatively complex and the amount of
computation is relatively large, there are still many open issues on
the details of our model, in particular, its deviation from steady
state, its stability, uniqueness and orbit composition.  While these
deserve to be addressed in a more systematic set of study, some
insights can already be obtained from the following simple analysis.

\section{Analysis and discussion}

\subsection{Mass fractions of orbit families in the bar}

Let us examine the orbit distribution of the final
model.  Figure~\ref{orbit-point.ps} shows the distributions of the
orbits in our model.  No chaotic orbits, retrograde orbits or orbits
without fixed sense of rotation are explicitly used to build the model
(see Section~\ref{lib}).  The direct regular orbits and the
collective-orbits of increasing weight are indicated with diamond and
plus symbols of increasing size.  The plus symbols trace a
one-parameter sequence of collective-orbits as a function of $E_J$
except for a gap between $E_J=-2.5$ and $E_J=-2.2$.  In the nucleus,
there are some collective-orbits and direct $x_2$ orbits ($y/x \sim 1$
and $E_J<-2.2$).  The direct banana orbits are in the region ($y/x
\sim 0.5$, $z/R>0.6$).  The axis ratio of the Dwek et al. model
would be at the coordinate ($y/x=0.4$, $z/R=0.25$) in the middle
panel.  The lower panel also shows the amount of time-dependency at
the end of the integration, which is less than 1\% after 100
epicycles.  These orbits are certainly valid to be used to construct
steady state model.

To examine the role of different orbit families in the
model more quantitatively, we plot the cumulative fractions of various
types of orbits in the self-consistent model as a function of the
Jacobi energy of an orbit in Figure~\ref{orbit-type-eng.ps}.  One can
see that the mass is roughly equally divided between the direct
regular orbits and the collective-orbits.  About one-fifth of the
direct orbits belong to the banana (2:2:1) family, and very little
mass in $x_2$ orbits.  The rest are mostly $x_1$ orbits.  A
collective-orbit lumps all possible orbits with the same Jacobi
integral together, some are regular, most are chaotic.  We estimate
based on typical surface-of-sections for a bar potential (see, e.g.,
Figure 13 in Sellwood and Wilkinson 1993) that about $2/3$ of mass in
our collective-orbits is on chaotic orbits, and about $1/3$ on
retrograde $x_4$ orbits; the direct orbits occupy very little phase
space of the collective-orbits.  In some sense collective-orbits are
alternative representations of (at least some of) retrograde orbits
and chaotic orbits.  The mass in collective-orbits implies about
$15\%$ of the bar's mass in the retrograde orbits and $30\%$ in the
chaotic orbits.  When we run other simulations without
collective-orbits but with explicit retrograde orbits and chaotic
orbits, we find similar fractions (Zhao 1994).

The large fraction of collective-orbits in our model ($\sim 45\%$)
comes from self-consistency on the minor axis.  A model with regular
orbits only would have an intrinsically peanut shape system, which
might also be peanut-shaped in projection.  But neither the volume
density model of Dwek et al. nor the dust corrected {\sl COBE} map
have peanut shape.  The chaotic orbits that are lumped in
collective-orbits can reach the minor axis.  The large fraction of
chaotic orbit is further enhanced by the central nucleus in the model.
A stationary nearly prolate bar with a finite core and an axis ratio
$1.9:1:0.7$ would have many box orbits (Schwarzschild 1979).
Apparently virtually all these orbits are destablized by the nucleus
in our bar.  Most of them may end up in chaotic orbits and some in the
retrograde $x_4$ orbits.

The significant fraction of retrograde orbits ($15\%$) and chaotic
orbits ($30\%$) in our model, as implied by collective-orbits, may
shed some light on the formation history of the bar.  It is a working
scenario that a bar develops from instability of a thin disk.  It
bends out of the plane a few rotations later and resymmetrizes to a
thickened bulge.  According to the KAM theory (e.g., Moser 1983),
regular orbits far from the resonances of an integrable system can
survive adiabatic changes in potential.  As virtually all disk stars
(except those in the rare counter-rotating discs) have direct sense of
rotation by definition, the retrograde or chaotic orbits in the
developed bar must come from scattering by the resonances and the
rapid fluctuations in potential in the bending phase.  As away from
the plane where gravity is weak many orbits scattered there can easily
switch their sense of rotation, it is likely that stars populate the
retrograde and chaotic orbits during this rapid bending phase.  It
would be interesting to examine how non-direct orbits form in N-body
simulations and test if their fraction can be used as an indicator of
the strength of relaxation in the bending phase.

\subsection{Residual, equilibrium and stability}

Although it fits observation quite well, our bar model has not yet
reached mathematical self-consistency; the quadrature sum residual of
the density is still at 0.5\% level.  Although a mathematically
self-consistent model could be a violently unstable system with very
noisy phase space, which nature would never make, a model with the
smaller residual is closer to (a stable or unstable) equilibrium.
Comparing with simple axisymmetric systems or stationary triaxial
systems, the orbit structure of the bar is more complex.  So the small
residual in our orbit model may reflect some level of incompleteness
in our orbit library, which consists of relatively small numbers of
regular direct orbits and collective-orbits.  However, it is also
unclear whether the Dwek et al. 3D volume density model, obtained from
a parametrized fit to the 2D {\sl COBE} map, have small unphysical
regions, which intrinsicly have no orbit counterparts.  Since the
input parametrized model still has systematic residuals in fitting the
{\sl COBE} light distribution (see Figure 3 of Dwek et al. and our
Figure~\ref{dered.ps}), there are even less reasons to believe that
the model has to be realizable to every detail.  More meaningful
deprojected model should use orbits as basis functions.

What are the effects of the residual?  It implies that the model will
evolve with time.  The small residual probably makes no difference in
the model's ability to predict and match observations if the model is
stable.  One needs to know how far the model will deviate from its
initial state and how stable it is.

To answer some of these questions, we convert our orbit model into an
N-body model and study how the system evolves.  The N-body model is
generated by sampling the weighted orbits in our steady state model
at random phase of the integration.  The particles are assigned equal
masses.  The N-body simulation is run with the Self-Consistent Field
method (Hernquist and Ostriker 1992).  We evolve N=30K particles with
a time step of 1 million years ($1/100$ of the rotation period).  The
softening of gravity is done by truncating the SCF expansion up to
quantum numbers $n=6$ and $l=4$.  In computing the gravity, we allow
the odd part of the Spherical Harmonics expansion to contribute.  A
fixed MN-disk potential is also included.

The N-body simulation shows that the residual is not significant to
cause disruption of the bar nor strong dynamical evolution.  The
system relaxes to a configuration close to the initial one in one
rotation (0.1 Gyr) and remains in nearly steady state for at least
another 9 rotations.  Figure~\ref{nbody.ps} shows snap shots of the
model at $t=0$ and $t=1$ Gyr.  The overall shape and density at the
two different times are similar except that the final state is
somewhat rounder than the initial state; the axis ratio changed from
$1:0.6:0.4$ to $1:0.7:0.4$.  A more detailed look at the variation is
shown in Figure~\ref{nbody1.ps}.  It is well-known that equilibrium
systems satisfy the (steady state) Virial theorem, namely $W+2K=0$,
where $W$ is the Claussius Virial of the system, and $K$ is the total
kinetic energy of the system.  This allows us to measure how close the
model is in equilibrium by computing $-2K/W$.  Figure~\ref{nbody1.ps}
shows $-2K/W$ vs $t$ for our N-body model.  At $t=0$, $-2K/W=0.98$ for
the model.  It stays close to unity for the next ten rotations.  The
initial deviation from unity is because the potential is slightly
different from what we used in orbit calculation.  The figure also
shows the moments of inertia of the bar as a function of time.  The
axis ratio of the bar $(I_{xx}:I_{yy}:I_{zz})^{\half}$, measured by
its moments of inertia along the three principal axes, settles to a
constant value $1:0.7:0.4$ after some oscillation in the first
rotation period.  One can see that the cross term $I_{XY}$ measured in
the rest frame follows closely to a sinusoidal curve of a rotating bar
with a constant amplitude and constant period, which is half of the
bar's period ${2\pi \over \Omega} \sim 0.1$ Gyrs.  A more quantitative
inspection of the pattern speed finds that it is a slow declining
function of time with a relatively large oscillation in the first
rotation period.

It is interesting to compare our N-body experiment for the Galactic
bar with the experiments for Schwarzschild's triaxial galaxy model by
Smith and Miller (1982).  Their simulations were carried out for 100K
particles for about 6 dynamical time scales in both
non-self-gravitating and self-gravitating conditions.  The major
semi-axis typically increased by 20\% in the inner region and dropped by
5\% in the outer region in the initial a quarter of a dynamical time,
followed then by a lasting gentle contraction at all radii till the
end of the run (see their Figure 5).  Overall the major semi-axis
shortened by 20\% while the intermediate and the minor semi-axies
appeared to have little evolution (see their Figure 4).  Based on
these, they claimed that Schwarzschild's model, which was in rigorous
equilibrium by design, was robust without growing disturbance of more
than 0.5 per crossing time.  Comparing with their experiments,
although our bar is not designed as self-consistently as the
Schwarzschild model, it has similar amplitudes of the initial
oscillations and seems to have settled more quickly to
quasi-equilibrium (in one rotation).  The strongest evolution comes
from the pattern speed, a unique property of bars, which has a secular
decline still less than 10\% per Gyr.  While the long term stability
and interactions with a live disk and halo
remain to be investigated, the bar made from our orbit model is in a
stable quasi-equilibrium.

These findings are perhaps not surprising.  Firstly the small residual
and the low level of time-dependency of the regular orbits in the
model do not suggest any rapid dynamical evolution of the model.
Secondly the model's phase space is constrained by matching a likely
stable systems made by nature.  Both the thickness of the bar
(Figure~\ref{nbody.ps}) and the absence of counter rotation, strong
anisotropy and sharp variations in the velocity distribution
(Figure~\ref{surf.ps} and~\ref{minormajor.ps}), also do not argue for
the bending instability of a thin bar with axis ratio more extreme
than 1:3 and/or velocity dispersion ratio more extreme than 0.6
(Merritt and Sellwood 1994).

Somewhat surprising is that the included nucleus does not destroy the
bar while the opposite has often been argued (e.g., Hasan et
al. 1993).  Unlike the box orbits, the $x_1$ orbits do not pass very
close to the center.  About 5-10\% of the bar's mass are enclosed
inside 0.5 kpc, about the radius of the inner Lindblad resonance,
which may not be sufficient to destruct the bar.  Also a major
difference with previous models is that both our bar and the nucleus
are kept self-consistent.

\vfill

\subsection{Uniqueness}

Another important question is how many different models can be built
to match the same observations with similar amount of residual.  While
the use of small amount of smoothing and the positivity constraint
make the numerical deprojection process stable to pixel-to-pixel
variations, it is unclear whether significantly different models can
satisfy the same constraints equally well and what further constraints
one can impose to distinguish models.  As the {\sl COBE} map only
constrains a 2-D distribution of the light, there is a wide of range
of compatible potentials with different bar mass, orientation and
pattern speed.  Also for models with the same potential, the orbit
composition can be non-unique.  To fully investigate the
non-uniqueness of models for the COBE bar, one needs to go through the
exercise in the previous sections for all compatible potentials and
search for models with different mass distribution as well as velocity
distribution.  Due to the complexity of the problem and the amount of
calculations involved, we will delay this issue to further studies.
Some preliminary results can be found in Zhao (1994), where models
with or without a disk or nucleus and models with or without direct or
retrograde orbits are investigated.  Quite certainly, one can say that
the {\sl COBE} map itself plus a mass-to-light ratio does not
constrain the model uniquely.  But with detailed stellar velocity
data, gas kinematics, and the microlensing data as well as
self-consistency and stability, one can hopefully limit the parameters
of the bar to a narrow range.

\section{ Conclusion and future work}\label{concl}

We have built a 3D steady state dynamical model for the Galactic
bar using a generalized Schwarzschild technique.  325 regular direct
boxy orbits are integrated in a rapidly rotating bar potential with
corotation at 3.3 kpc, which includes a Miyamoto-Nagai disk, the Dwek
et al. bar and the $r^{-1.85}$ nucleus.  These orbits make up 55\% of
the bar's mass.  The rest of the mass is distributed according to an
implicit distribution function $f_c(E_J)$, which is numerically
divided into 160 independent single energy building blocks.  The mass
on these orbits and these building blocks are determined with the NNLS
method to fit the observations, keeping a relatively smooth
distribution in the phase space.

The model fits Dwek et al.'s luminosity model for the {\sl
COBE} bulge (see Figures~\ref{dered.ps},~\ref{den.ps} and
{}~\ref{surf.ps}) and existing kinematic data.  In particular, the orbit
model fits the velocity dispersion at Baade's window and the observed
solid body rotation curve of the bulge tracers.  The observed fall-off
of radial dispersion along the minor axis, the vertex deviation and
the proper motion anisotropy at Baade's window follow naturally from
the model (see Figure~\ref{minormajor.ps}).  The model is also in
agreement with the flat gas rotation curve of the inner Galaxy and the
asymmetry of light seen in the {\sl COBE} infrared maps of the bulge.

Our bar model is dominated by the regular direct boxy orbits with
about 15\% of bar's mass in retrograde orbits and 30\% in chaotic
orbits.  There are 10\% banana (2:2:1) orbits.

Following it with N-body simulations, we find the model is stable in
spite of the nucleus.  We conclude that the model is qualified to
interpret observations and a table for the predicted velocity and
dispersion across the bulge is given.

The steady state bar models have many potential astronomical
applications.  Up to now the MACHO team and the OGLE team have
together obtained more than 100 microlensing events in several fields
of the bulge.  To interpret these data, one needs a model that can
deliver information about the proper motion velocity and distance
distributions of both the lens and the source for the whole bulge.
The steady state model here is well suited for this purpose, as
first shown by Zhao et al. (1995).  This provides the only way to
probe the lower end of the mass function of the bar.

The technique here can be used to set up nearly equilibrium initial
conditions for N-body studies of stability and secular evolution.  As
most systems do not have analytical distribution functions, it is
difficult to set up an initially equilibrium model.  With the
Schwarzschild technique, one can in principal set up the full range of
equilibria, including theoretically interesting triaxial or
axisymmetric models with three integrals of motion and models that fit
particular observations.  One needs not to be limited to
axisymmetric systems with a closed set of moment equations, or the
narrow range of bars developed in a previous N-body simulation of an
unstable disk.  The wide range of initial states are better suited to
address the stability, the life span and the range of bars.  Most
relevant to the formation and evolution of the Galaxy is the stability
of the observationally well-constrained Galactic bar and the response
of the disk and the halo (Hernquist and Weinberg 1992).

Looking beyond the {\sl COBE} bar, steady state models can have
important applications in ellipticals, extragalactic bulges and
nuclei, which share the basic dynamics with the Galactic bar.  There
are plenty of evidences that suggest massive central black holes in
nearby galactic nuclei, including the recent finding of many cusped
nuclei by the Hubble Space Telescope (Lauer et al. 1993).  But the
basic ambiguity in dynamically constraining the black hole mass has
been the unknown amount of radial anisotropy, which can masquerade as
the gravity of a dark component.  Previous models often make use of
some of the following simplifying assumptions to keep calculations
tractable: 1) the potential is spherical (Dressler and Richstone
1988), or at least axisymmetric, 2) the intrinsic velocity has an
isotropic Gaussian distribution, or at least a distribution with two
integrals of motion only (Qian et al. 1995).  Although some of these
models already fit the observations remarkablely well (e.g., the
$f(E,L_z)$ models for the M32 nucleus by Qian et al. 1995 and Dehnen
1995) and are worth to be tested in other systems, since there is no
compelling reason to believe that real galactic nuclei satisfy the
above assumptions, it is eventually necessary to search the range of
the solution space in triaxial models or at least axisymmetric models
with three integrals of motion.  The available high quality data and
the new techniques of deriving velocity profiles (Rix and White 1992,
Gerhard 1993, van der Marel and Franx 1993) should give theorists
additional incentives to explore beyond a few relatively
easy-to-compute models.

Our orbit construction program is well suited to these
systems, as it works with the minimal assumptions of self-consistency,
positivity and smoothness.  It is the more {\sl proper technique} than
solving truncated moment equations, as one can actually {\sl fit} the
kinematic data by adjusting the weights of the orbits.  This unique
property allows one to make the complete use of the data, which now
includes skewness and the kurtosis of the line profile as well as the
streaming velocity, dispersion and surface brightness distribution at
subarcsec seeing at many positions of the nucleus.

In the next few years, there will be large surveys of stellar proper
motion in the bulge, e.g., for the OH/IR stars in the central cluster
from the group in Leiden (de Zeeuw 1993) and for the infrared bright
sources in the central pc from the group in Garching (Genzel 1995).
Combined with radial velocity data, the proper motions will yield the
3D velocity ellipsoid, which can place important constraints on the
triaxiality and the bar's orbit distribution (Zhao et al. 1994).  Most
of Spaenhauer et al.'s (1992) proper motion stars on Baade's window
are also measured for radial velocities (Terndrup et al. 1995a).  Our
steady state model shows a first step in linking these kinematic
data together with the {\sl COBE} map.


I would like to thank David N. Spergel and R. Michael Rich for guiding
me and contributing many ideas throughout this work, David Merritt,
Hans-Walter Rix and Nicolas Cretton for several enlightening
discussions on modelling technique, Jerry Sellwood, Tim de Zeeuw,
Simon White, Ortwin Gerhard and the referee James Binney for valuable
comments on an earlier version.  This work is a major outgrowth of PhD
thesis done at Columbia University, and is supported by Long-Term
Space Astrophysics grant NAGW-2479 to R. Michael Rich and NSF grant
AST 91-17388 and NASA grant ADP NAG5-2693 to David N. Spergel.



\begin{figure}[h]
\vbox to4in{\rule{0pt}{4in}}
\caption{ The upper panel shows the distribution of the orbits in the
energy vs angular momentum plane.  The middle panel shows the axis
ratio distribution.  The lower panel shows how well the regular orbit
fills its 3D torus at the end of integration.  Each isolated dot
indicates an orbit assigned very low non-zero weight, dots with
diamond and plus symbols of increasing size indicate the regular
direct boxy orbits and collective-orbits of the steady state
bar with increasing weights.  The dashed lines in the upper panel
correspond to the dynamical boundaries.  Most orbits are direct
orbits, supporting the bar.  }\label{orbit-point.ps}
\end{figure}

\begin{figure}[h]
\vbox to2in{\rule{0pt}{2in}}
\caption{ plots the projected maps of the bar
dynamical model (solid line) and the modified Dwek volume density
model (dashed line) over a dust-subtracted K band contour map of the
inner Galaxy from {\sl COBE} (dotted line); contours are spaced with
one magnitude interval.  For the models, a Miyamoto-Nagai disk is
included for direct comparison with the {\sl COBE} map.
}\label{dered.ps}
\end{figure}

\begin{figure}[h]
\vbox to3in{\rule{0pt}{3in}}
\caption{ shows the circular speed of the model as a function of
radius along the intermediate $y$ axis of the bar.  The model
potential is from a modified Dwek bar plus a Miyamoto-Nagai disk, with
the bar potential computed from the first 10 (solid line) and 84
(dashed line) terms of the Hernquist-Ostriker expansions.  The halo
potential is neglected for the inner 3 kpc.  Note Different
truncations give only less than 20 km/s difference in velocity and
both produce the flat rotation curve, characteristic of the Galaxy.
}\label{rotcurve.ps}
\end{figure}

\begin{figure}[h]
\vbox to4in{\rule{0pt}{4in}}
\caption{ compares the mass distribution of the orbits (solid line)
with the volume density of the input modified Dwek model (dashed line)
in the yz and xy planes.  Contours are spaced with a factor of 2
interval.  The residual between the two models is only 0.5\% in
quadratures of density.  }\label{den.ps}
\end{figure}

\begin{figure}[h]
\vbox to4in{\rule{0pt}{4in}}
\caption{
{}From top to bottom the solid contours are the surface density,
line-of-sight velocity and dispersion maps of the bar
model in the Galactic coordinates.  Also shown in dashed contours is
surface density of the modified Dwek et. al. model, spaced with one
magnitude interval.  Note the model fits the boxyness and the
asymmetries of the input density model up to $10^o$ from the center,
it and predicts a solid body rotation field and radial fall off of the
dispersion.  }\label{surf.ps}
\end{figure}

\begin{figure}[h]
\vbox to4in{\rule{0pt}{4in}}
\caption{  The upper panel plots the
radial velocity dispersion ($\sigma_r$) and rotation velocity ($V_r$)
along the $b=-4^o$ slice, the $b=-7^o$ slice from the model.  The
lower panel plots minor axis runs of the radial velocity dispersion
for the model (solid line) and observations (diamond symbols), the
longitude and latitude proper motion dispersion $\sigma_l$ and
$\sigma_b$ and the cross term $\sigma_{lr}$ and its measurement at
Baade's Window (asterisk).  Note the fall-off of dispersion on the
minor axis.  See text for the references to the data points.
}\label{minormajor.ps}
\end{figure}

\begin{figure}[h]
\vbox to4in{\rule{0pt}{4in}}
\caption{shows the fractions of various types of orbits in the
model with the energy lower than than $E_J$.  The
Jacobi energy $E_J$ measures the radial extent of an orbit; at
$E_J=-2.2$ and $-1.8$, the typical radius of an orbit is $0.5$ kpc and
$1$ kpc respectively.  Most of the mass is in the regular direct boxy
orbits. }\label{orbit-type-eng.ps}
\end{figure}

\begin{figure}[h]
\vbox to3in{\rule{0pt}{3in}}
\caption{ plots an N-body realization of the steady state model (the
left panels), and the configuration after evolving for 10 rotation
periods (the right panels).  The solid line indicates our line-of-sight
to the center.  Note the elongated bar shape in the face-on view (the
lower panels) and the boxyness in the edge-on view (the upper panels)
are similar at two epochs.  The final bar has settled down to
dynamical equilibrium.  }\label{nbody.ps}
\end{figure}

\begin{figure}[h]
\vbox to4in{\rule{0pt}{4in}}
\caption{shows the time evolution of some global indicators of the N-body
bar in the upper panel the $-2K/W$ and the renormalized pattern speed.
(should all be unity if in steady state), and in the lower panel the
three moments of inertia $I_{xx}$, $I_{yy}$ and $I_{zz}$ along the
three principal axes (should all be constant) and the rest frame cross
term $I_{XY}$ (should be sinusoidal).  }\label{nbody1.ps}
\end{figure}

\onecolumn
\vfill\eject

 \begin{planotable}{ lrrrrr }
 \tablewidth{33pc}
 \tablecaption
 {Predicted distributions of the line-of-sight velocity $V_r$
and dispersion $(\sigma_r)$ over the inner $10^o$ of
the Galactic bulge}
 \tablehead{
\colhead{$V_r(\sigma_r)$} &\colhead{$l=\pm 1^o$}
&\colhead{$l=\pm 3^o$} &\colhead{$l=\pm 5^o$}
&\colhead{$l=\pm 7^o$} &\colhead{$l=\pm 9^o$}
 }
 \startdata
$b= 0^o$&  24(117)&  42(134)&  74(118)&  88(107)&  97( 88)
 \nl
        & -25(117)& -43(132)& -80(115)&-101(100)&-113( 88)
 \nl
$b= 1^o$&  27(131)&  45(133)&  76(118)&  96(105)& 103( 87)
 \nl
        & -28(129)& -47(131)& -83(113)&-107( 99)&-119( 86)
 \nl
$b= 2^o$&  37(130)&  55(126)&  78(112)&  99(101)& 106( 88)
 \nl
        & -39(128)& -61(122)& -88(109)&-115( 95)&-125( 82)
 \nl
$b= 3^o$&  20(122)&  51(120)&  76(104)& 100( 96)& 108( 84)
 \nl
        & -23(120)& -58(115)& -92(100)&-120( 89)&-130( 79)
 \nl
$b= 4^o$&  15(113)&  52(110)&  70( 97)&  92( 86)& 117( 77)
 \nl
        & -25(111)& -64(108)& -94( 93)&-112( 80)&-128( 76)
 \nl
$b= 5^o$&  12( 98)&  44( 98)&  68( 93)&  87( 74)& 108( 68)
 \nl
        & -22(102)& -61( 95)& -96( 83)&-101( 77)&-121( 78)
 \nl
$b= 6^o$&   9( 84)&  30( 84)&  66( 82)&  88( 68)&  97( 60)
 \nl
        & -30( 81)& -60( 78)& -90( 76)& -86( 79)& -97( 82)
 \nl
$b= 7^o$&   0( 72)&  30( 74)&  62( 75)&  84( 62)&  94( 57)
 \nl
        & -36( 69)& -65( 69)& -77( 81)& -68( 80)& -91( 69)
 \nl
$b= 8^o$&   3( 67)&  36( 64)&  76( 74)&  79( 57)&  82( 57)
 \nl
        & -31( 69)& -58( 84)&-100( 88)&-113( 60)& -92( 67)
 \nl
$b= 9^o$&  22( 83)&  43( 67)&  86( 63)& 103( 60)&  71( 32)
 \nl
        & -35( 83)& -57( 75)& -94( 79)& -98( 66)& -65( 75)
 \nl
$b=10^o$&  14( 63)&  33( 60)&  58( 64)&  90( 59)&  68( 41)
 \nl
        & -43( 71)& -53( 58)& -81( 65)& -98( 75)& -50( 61)
 \nl
 \tablecomments
 {At each latitude $b$, the upper and lower rows are for
the positive and negative longitude $l$ fields respectively.
Both $V_r$ and $\sigma_r$ are in units of km/s and in
Galactocentric frame; the latter is bracketed in the table.}
 \end{planotable}

 \begin{planotable}{ lrrrrr }
 \tablewidth{40pc}
 \tablecaption
 {Predicted distributions of the cross term $\sigma_{lr}$
and proper motion dispersions $[\sigma_l,\sigma_b]$ for the Galactic bulge}
 \tablehead{
\colhead{$\sigma_{lr}[\sigma_l,\sigma_b]$}
&\colhead{$l=\pm 1^o$} &\colhead{$l=\pm 3^o$}
&\colhead{$l=\pm 5^o$} &\colhead{$l=\pm 7^o$} &\colhead{$l=\pm 9^o$}
 }
 \startdata
$b= 0^o$&-34[133 127]&-41[134 132]&-51[127 131]&-46[116 124]&-41[101 119]
 \nl
        &-31[133 127]&-36[136 132]&-38[128 128]&-22[119 124]& 19[110 117]
 \nl
$b= 1^o$&-45[143 134]&-47[134 129]&-50[125 128]&-46[115 122]&-42[103 117]
 \nl
        &-44[143 134]&-40[135 128]&-37[128 124]&-20[118 120]& 21[109 114]
 \nl
$b= 2^o$&-57[141 122]&-55[135 120]&-52[122 119]&-49[115 115]&-46[ 97 109]
 \nl
        &-53[140 123]&-46[136 119]&-29[128 116]&-25[114 111]& 18[105 107]
 \nl
$b= 3^o$&-59[146 118]&-55[134 112]&-47[121 111]&-44[112 106]&-45[ 97 102]
 \nl
        &-57[148 117]&-40[137 110]&-23[128 105]&  8[112 100]& 31[ 96  99]
 \nl
$b= 4^o$&-58[140 106]&-54[133 102]&-36[116 103]&-36[105  97]&-41[ 92  89]
 \nl
        &-57[141 106]&-38[135 103]&-23[125 100]& 17[109  93]& 43[ 91  93]
 \nl
$b= 5^o$&-53[141  97]&-57[130  94]&-35[114  94]&-37[ 91  89]&-37[ 87  83]
 \nl
        &-46[137  97]&-40[129  93]&-25[115  88]& 45[105  91]& 46[103  90]
 \nl
$b= 6^o$&-53[120  92]&-39[121  89]&-33[ 97  90]&-36[ 84  85]&-32[ 90  86]
 \nl
        &-45[120  93]&-42[118  86]& 10[118  85]& 61[121  89]& 47[113  88]
 \nl
$b= 7^o$&-29[129  81]&-46[114  81]&-39[ 94  88]&-31[ 81  83]&-32[ 85  80]
 \nl
        &-35[136  82]&-24[138  78]& 56[143  88]& 69[134  92]& 38[ 91  77]
 \nl
$b= 8^o$&-36[116  78]&-39[104  79]&-53[ 99  80]&-36[ 78  85]&-23[ 69  72]
 \nl
        &-15[142  81]& 57[151  79]& 67[125  78]& 45[ 98  86]& 38[101  79]
 \nl
$b= 9^o$&-12[121  81]&-40[105  86]&-50[ 86  86]&-48[ 71  83]&-23[ 53  71]
 \nl
        & 50[133  87]& 41[123  80]&  8[ 96  71]&-16[ 85  76]&-45[ 74  74]
 \nl
$b=10^o$&-46[114  84]&-41[ 80  94]&-34[ 75  95]&-39[ 53  82]&-16[ 51  72]
 \nl
        &-20[ 98  85]&-28[121  75]&-41[119  61]&-25[ 77  67]&-20[ 51  70]
 \nl
 \tablecomments
 {See the note for Table 1.}
 \end{planotable}

\end{document}